\documentclass[11pt, oneside]{article}

\usepackage{amssymb}
\usepackage{amsmath}
\usepackage{amsthm}
\usepackage{amstext}

\theoremstyle{plain}
\newtheorem{theorem}{Theorem}[section]
\newtheorem{lemma}[theorem]{Lemma}
\newtheorem{corollary}[theorem]{Corollary}
\theoremstyle{definition}
\newtheorem{definition}[theorem]{Definition}

\newcommand*{\tu}{two-universal}
\newcommand*{\tuty}{\tu ity}

\newcommand*{\bbN}{\mathbb{N}}

\newcommand*{\cB}{\mathcal{B}} 
\newcommand*{\cE}{\mathcal{E}}
\newcommand*{\cF}{\mathcal{F}}

\newcommand*{\cH}{\mathcal{H}}

\newcommand*{\cS}{\mathcal{S}}

\newcommand*{\cX}{\mathcal{X}}
\newcommand*{\cY}{\mathcal{Y}}
\newcommand*{\cZ}{\mathcal{Z}}

\newcommand*{\epsb}{\bar{\varepsilon}}

\newcommand*{\event}{\mathcal{E}} 

\newcommand*{\Hq}{S}

\newcommand*{\ket}[1]{| #1 \rangle}
\newcommand*{\bra}[1]{\langle #1 |}

\newcommand*{\tr}{\mathrm{tr}}
\newcommand*{\rank}{\mathrm{rank}}

\newcommand*{\rvrho}{\boldsymbol{\rho}}
\newcommand*{\rvsigma}{\boldsymbol{\sigma}}
\newcommand*{\proj}[1]{\ket{#1} \bra{#1}}
\newcommand*{\dist}{\delta}

\newcommand*{\dd}{d}

\newcommand*{\mydtwo}{D}
\newcommand*{\mydeltatwo}{\Delta}

\title{{\bf Universally Composable Privacy Amplification Against
    Quantum Adversaries}\footnote{This work was partially supported by
    the Swiss National Science Foundation, project No.~20-66716.01.}}

\author{Renato Renner \quad Robert K\"onig \vspace{2ex} \\
             Computer Science Department \\
             ETH Z\"urich;
             Switzerland \\
             {\tt renner@inf.ethz.ch} \quad {\tt rkoenig@inf.ethz.ch}}

\date{}

\begin{document}
\maketitle

\begin{abstract}
  Privacy amplification is the art of shrinking a partially secret
  string $Z$ to a highly secret key $S$. We show that, even if an
  adversary holds quantum information about the initial string $Z$,
  the key $S$ obtained by two-universal hashing is secure, according
  to a universally composable security definition.  Additionally, we
  give an asymptotically optimal lower bound on the length of the
  extractable key $S$ in terms of the adversary's (quantum) knowledge
  about $Z$. Our result has applications in quantum cryptography.  In
  particular, it implies that many of the known quantum key
  distribution protocols are universally composable.
\end{abstract}

\section{Introduction}

\subsection{Privacy amplification}

Consider two parties having access to a common string $Z$ about which
an adversary might have some partial information.  \emph{Privacy
  amplification}, introduced by Bennett, Brassard, and
Robert~\cite{BeBrRo88}, is the art of transforming this partially
secure string $Z$ into a highly secret key $S$ by public discussion.
A good technique is to compute $S$ as the output of a publicly chosen
\tu{} hash function\footnote{See Section~\ref{sec:tudef} for a
  definition of \tu{} functions.} $F$ applied to $Z$. Indeed, it has
been shown \cite{BeBrRo88,ImLelu89,BBCM95} that, if the adversary
holds purely classical information $W$ about $Z$, this method yields a
secure key $S$ and, additionally, is asymptotically optimal with
respect to the length of $S$.  For instance, if both the initial
string $Z$ and the adversary's knowledge $W$ consist of many
independent and identically distributed parts, the number of
extractable key bits roughly equals the conditional Shannon entropy
$H(Z|W)$.

The analysis of privacy amplification can be extended to a situation
where the adversary might hold quantum instead of only classical
information about $Z$.  It has been shown~\cite{KoMaRe03} that \tu{}
hashing allows for the extraction of a secure key $S$ whose length
roughly equals the difference between the entropy of $Z$ and the
number of qubits stored by the adversary. This can be applied to
proving the security of quantum key distribution (QKD) protocols where
privacy amplification is used for the classical post-processing of the
(only partially secure) raw key \cite{ChReEk04}.

\subsection{Universal composability}


Cryptographic protocols (e.g., for generating a secret key) are often
used as components within a larger system (where, e.g., the secret key
is used to encrypt messages). It is thus natural to require that the
security of a protocol is not compromised when it is, e.g., invoked as
a sub-protocol in any (arbitrarily complex) scheme. This requirement
is captured by the notion of universal composability.  Roughly
speaking, a cryptographic protocol is said to be \emph{universally
  composable} if it is secure in \emph{any} arbitrary context. For
instance, the universal composability of a secret key $S$ guarantees
that any bit of $S$ remains secret even if some other part of $S$ is
given to an adversary.\footnote{Note that this is not necessarily the
  case for many known security definitions of a secret key.}

In the past few years, composable security has attracted a lot of
interest and lead to important new definitions and proofs (see, e.g.,
the framework of Canetti~\cite{Canetti01} or Pfitzmann and
Waidner~\cite{PfiWai00a}).  Recently, Ben-Or and Mayers have
generalized the notion of universal composability to the quantum
case~\cite{BenMay02}. Universally composable security definitions are
usually based on the idea of characterizing the security of a
cryptographic scheme by its distance to an ideal system which (by
definition) is perfectly secure. For instance, a secret key $S$ is
universally composable if it is close to an independent and almost
uniformly distributed string $U$. This then implies that any
cryptosystem which is proven secure when using a perfect key $U$
remains secure when $U$ is replaced by the (real) key $S$.

Ben-Or, Horodecki, Leung, Mayers, and Oppenheim~\cite{BHLMO02} were
the first to address the problem of universal composability in the
context of QKD.  Usually, the security of a QKD scheme is defined by
the requirement that the mutual information between the final key $S$
and the outcome of an arbitrary measurement of the adversary's quantum
system be small (for a formal definition, see, e.g., \cite{NieChu00}
or \cite{GotLo03}). This, however, does not necessarily imply
composability. Indeed, an adversary might wait with the measurement of
his quantum state until he learns some of the bits of $S$, which might
allow him to obtain more information about the remaining bits.

\subsection{Contributions}

We analyze the security of privacy amplification in a setting where an
adversary holds quantum information. We show that the key obtained by
\tu{} hashing is secure according to a very strong security definition
which, in any context, guarantees virtually the same security as a
perfect key. The security definition we use is essentially equivalent
to the definition used in~\cite{BHLMO02} for analyzing the
composability of QKD, and thus also provides universal composability
with respect to the framework of~\cite{BenMay02} (cf.\ 
Section~\ref{sec:uc}). This extends the result of~\cite{KoMaRe03}
where a weaker (not necessarily composable) security definition has
been used. Moreover, our results have implications for quantum
cryptography. In particular, it follows from the analysis
in~\cite{ChReEk04} (which is based on the security of privacy
amplification) that many of the known QKD protocols (such as
BB84~\cite{BenBra84} or B92~\cite{Bennett92}) are universally
composable (cf.\ Section~\ref{sec:QKD} for more details).

Additionally, we improve the lower bound on the length of the
extractable key $S$ given in~\cite{KoMaRe03}. If the initial
information $Z$ as well as the adversary's (quantum) knowledge consist
of $n$ independent pieces, our bound is asymptotically tight, for $n$
approaching infinity. In particular, we obtain an explicit expression
(in terms of von Neumann entropy) for the rate at which secret key
bits can be generated, thus generalizing a result which has only been
known for the case of purely classical adversaries (cf.\ 
Section~\ref{sec:opt}).





\section{Preliminaries} \label{sec:pre}

\subsection{Random functions and \tu{} functions} \label{sec:tudef}

A \emph{random function} from $\cX$ to $\cY$ is a random variable
taking values from the set of functions with domain $\cX$ and range
$\cY$.  A random function $F$ from $\cX$ to $\cY$ is called
\emph{\tu{}} if
\[
  \Pr[F(x) = F(x')] \leq \frac{1}{|\cY|} \ ,
\]
for any distinct $x, x' \in \cX$.\footnote{In the literature, \tuty{}
  is usually defined for families $\cF$ of functions: A family $\cF$
  is called \tu{} if the random function $F$ with uniform distribution
  over $\cF$ is \tu.}  In particular, $F$ is \tu{} if, for any
distinct $x, x' \in \cX$, the random variables $F(x)$ and $F(x')$ are
independent and uniformly distributed. For instance, the random
function chosen uniformly from the set of all functions from $\cX$ to
$\cY$ is \tu.  Non-trivial examples of \tu{} functions can, e.g., be
found in~\cite{CarWeg79} and~\cite{WegCar81}.

\subsection{Density operators and random states}

Let $\cH$ be a Hilbert space. We denote by $\cS(\cH)$ the set of
\emph{density operators} on $\cH$, i.e., $\cS(\cH)$ is the set of
positive operators $\rho$ on $\cH$ with $\tr(\rho)=1$. A density
operator $\rho \in \cS(\cH)$ is called \emph{pure} if it has rank $1$,
i.e., $\rho = \proj{\phi}$ for some $\ket{\phi} \in \cH$.

Let $(\Omega, P)$ be a discrete probability space. A \emph{random
  state} $\rvrho$ on $\cH$ is a random variable with range $\cS(\cH)$,
i.e., a function from $\Omega$ to $\cS(\cH)$. Let $\rvrho$ and
$\rvrho'$ be two random states on $\cH$ and $\cH'$, respectively. The
\emph{tensor product} $\rvrho \otimes \rvrho'$ of $\rvrho$ and
$\rvrho'$ is the random state on $\cH \otimes \cH'$ defined by
\[
  (\rvrho \otimes \rvrho')(\omega) 
:=
  \rvrho(\omega) \otimes \rvrho'(\omega) \ ,
\]
for any $\omega \in \Omega$.

To describe settings involving both classical and quantum information,
it is often convenient to represent classical information as a state
of a quantum system. Let $X$ be a random variable with range $\cX$ and
let $\cH$ be a $|\cX|$-dimensional Hilbert space with orthonormal
basis $\{\ket{x}\}_{x \in \cX}$.  The \emph{random state
  representation} of $X$, denoted $\{X\}$, is the random state on
$\cH$ defined by $\{X\} := \proj{X}$, i.e., for any $\omega \in
\Omega$,
\[
  \{X\}(\omega) = \proj{X(\omega)} \ .
\]
Let $\rvrho$ be a random state. For an observer which is ignorant of
the randomness of $\rvrho$, the density operator of the quantum system
described by $\rvrho$ is given by
\[
  [\rvrho] 
:= 
  E_{\rvrho}[\rvrho] 
=
    \sum_{\omega \in \Omega} 
      P(\omega) \rvrho(\omega) \ .
\]
More generally, for any event $\event$, the \emph{density operator of}
$\rvrho$ \emph{conditioned on} $\event$, denoted $[\rvrho|\event]$, is
defined by
\[
  [\rvrho|\event] 
:= 
  E_{\rvrho}[\rvrho|\event] 
=
  \frac{1}{\Pr[\event]} 
    \sum_{\omega \in \event} 
      P(\omega) \rvrho(\omega) \ .
\]

Let $\rvrho \otimes \{X\}$ be a random state consisting of a classical
part $\{X\}$ specified by a random variable $X$. It is easy to see
that the corresponding density operator $[\rvrho \otimes \{X\}]$ is
given by
\begin{equation} \label{eq:combexp}
  [\rvrho \otimes \{X\}] 
=
  E_X\bigl[\rho_X \otimes \proj{X}\bigr]
\end{equation}
where $\rho_x := [\rvrho|X=x]$. In particular, if $X$ is independent
of $\rvrho$, then
\begin{equation} \label{eq:indepprod}
  [\rvrho \otimes \{X\}]  = [\rvrho] \otimes [\{X\}] \ .
\end{equation}

\subsection{Distance measures and non-uniformity}

The \emph{variational distance} between two probability distributions
$P$ and $Q$ over the same range $\cX$ is defined as
\[
\dist(P,Q) := \frac{1}{2} \sum_{x \in \cX} |P(x) - Q(x)| \ .
\]
The variational distance between two probability distributions $P$ and
$Q$ can be interpreted as the probability that two random experiments
described by $P$ and $Q$, respectively, are different.  This is
formalized by the following lemma.

\begin{lemma} \label{lem:vardistevent}
  Let $P$ and $Q$ be two probability distributions.  Then there exists
  a pair of random variables $X$ and $X'$ with joint probability
  distribution $P_{X X'}$ such that $P_X = P$, $P_{X'} = Q$, and
  \[
  \Pr[X \neq X'] = \dist(P, Q) \ .
  \]
\end{lemma}

The \emph{trace distance} between two density operators $\rho$ and
$\sigma$ on the same Hilbert space $\cH$ is defined as
\[
  \dist(\rho, \sigma) := \frac{1}{2} \tr(|\rho-\sigma|) \ .
\] 
The trace distance is a metric on the set of density operators
$\cS(\cH)$. We say that $\rho$ is \emph{$\varepsilon$-close} to
$\sigma$ if $\delta(\rho, \sigma) \leq \varepsilon$, and denote by
$\cB^{\varepsilon}(\rho)$ the set of density operators which are
$\varepsilon$-close to $\rho$, i.e., $\cB^{\varepsilon}(\rho) =
\{\sigma \in \cS(\cH): \dist(\rho, \sigma) \leq \varepsilon\}$.

The trace distance is subadditive with respect to the tensor product,
i.e., for any $\rho, \sigma \in \cS(\cH)$ and $\rho', \sigma' \in
\cS(\cH')$,
\begin{equation} \label{eq:trbnd}
  \dist(\rho \otimes \rho', \sigma \otimes \sigma') 
\leq
  \dist(\rho, \sigma) + \dist(\rho', \sigma') \ ,
\end{equation}
with equality if $\rho' = \sigma'$, i.e.,
\begin{equation} \label{eq:trident}
  \dist(\rho \otimes \rho', \sigma \otimes \rho')
=
  \dist(\rho, \sigma) \ .
\end{equation}
Moreover, it cannot increase when the same quantum operation $\cE$ is
applied to both arguments, i.e.,
\begin{equation} \label{eq:distop}
  \dist(\cE(\rho), \cE(\sigma)) \leq \dist(\rho, \sigma) \ .
\end{equation}
Similarly, the trace distance between $\rho$ and $\sigma$ is an upper
bound for the variational distance between the probability
distributions $P$ and $Q$ of the outcomes when applying the same
measurement to $\rho$ and $\sigma$, respectively, i.e.,
\begin{equation} \label{eq:distmeas}
  \dist(P, Q) \leq \dist(\rho, \sigma) \ .
\end{equation}

The variational distance can be seen as a (classical) special case of
the trace distance. Let $X$ and $Y$ be random variables. Then the
variational distance between the probability distributions of $X$ and
$Y$ equals the trace distance between the corresponding density
matrices $[\{X\}]$ and $[\{Y\}]$, i.e.,
\[
  \dist(P_X, P_Y) = \dist([\{X\}], [\{Y\}]) \ .
\]

The trace distance between two density operators containing a
representation of the same classical random variable $X$ can be
written as the expectation of the trace distance between the density
operators conditioned on $X$.

\begin{lemma} \label{lem:distexp}
  Let $X$ be a random variable and let $\rvrho$ and $\rvsigma$ be
  random states. Then
  \[
  \dist([\rvrho \otimes \{X\}], [\rvsigma \otimes \{X\}])
  =
  E_X[\dist(\rho_X, \sigma_X)]
  \]
  where $\rho_x:=[\rvrho|X=x]$ and $\sigma_x:=[\rvsigma|X=x])$.
\end{lemma}

\begin{proof}
  Using~(\ref{eq:combexp}) and the orthogonality of the vectors
  $\ket{x}$, we obtain
  \[
  \begin{split}
    \dist([\rvrho \otimes \{X\}], [\rvsigma \otimes \{X\}])
  =
    \frac{1}{2} \tr \Bigl|
      E_X\bigl[(\rho_X-\sigma_X) \otimes \proj{X} \bigr]
    \Bigr| 
  =
    \frac{1}{2} \tr \Bigl(
      E_X \Bigl[
        \bigl| (\rho_X - \sigma_X) \otimes \proj{X} \bigr| 
      \Bigr] \Bigr) \ .
  \end{split}
  \]
  The assertion then follows from the linearity of the trace and the
  fact that $\tr \bigl| (\rho_x - \sigma_x) \otimes \proj{x}\bigr| =
  \tr |\rho_x - \sigma_x|$.
\end{proof}

In Section~\ref{sec:uc}, we will see that a natural measure for
characterizing the secrecy of a key is its trace distance to a uniform
distribution.

\begin{definition}
  Let $X$ be a random variable with range $\cX$ and let $\rvrho$ be a
  random state. The \emph{non-uniformity} of $X$ \emph{given} $\rvrho$
  is defined by
  \[
    \dd(X|\rvrho) 
  := 
    \dist([\{X\} \otimes \rvrho], [\{U\}] \otimes [\rvrho])
  \]
  where $U$ is a random variable uniformly distributed on $\cX$.  
\end{definition}

Note that $\dd(X|\rvrho) = 0$ if and only if $X$ is uniformly
distributed and independent of $\rvrho$.

\subsection{(Smooth) R\'enyi entropy}

Let $\rho \in \cS(\cH)$ be a density operator and let $\alpha \in [0,
\infty]$. The \emph{R\'enyi entropy of order $\alpha$ of $\rho$} is
defined by
\[
  \Hq_{\alpha}(\rho)
:=
  \frac{1}{1-\alpha} \log_2 \bigl( \tr(\rho^\alpha) \bigr)
\]
with the convention $\Hq_{\alpha}(\rho) := \lim_{\beta \to \alpha}
\Hq_{\beta}(\rho) $ for $\alpha \in \{0, 1, \infty\}$. In particular,
for $\alpha = 0$, $\Hq_{0}(\rho) = \log_2\bigl(\rank(\rho)\bigr)$ and,
for $\alpha = \infty$, $\Hq_{\infty}(\rho) = -
\log_2\bigl(\lambda_{\max}(\rho)\bigr)$ where $\lambda_{\max}(\rho)$
denotes the maximum eigenvalue of $\rho$. For $\alpha=1$,
$\Hq_{\alpha}(\rho)$ is equal to the von Neumann entropy $\Hq(\rho)$.
Moreover, for $\alpha, \beta \in [0, \infty]$,
\begin{equation} \label{eq:Halpharel}
  \alpha \leq \beta \iff \Hq_{\alpha}(\rho) \geq \Hq_{\beta}(\rho) \ .
\end{equation}
Note that, for a classical random variable $X$, the R\'enyi entropy
$\Hq_{\alpha}([\{X\}])$ of the quantum representation of $X$
corresponds to the R\'enyi entropy $H_{\alpha}(X)$ of $X$ as defined
in classical information theory~\cite{Renyi61}.

The definition of R\'enyi entropy for density operators can be
generalized to the notion of smooth R\'enyi entropy, which has been
introduced in~\cite{RenWol03u} for the case of classical probability
distributions.

\begin{definition}
  Let $\rho \in \cS(\cH)$, let $\alpha \in [0,\infty]$, and let
  $\varepsilon \geq 0$.  The \emph{$\varepsilon$-smooth R\'enyi
    entropy of order $\alpha$ of $\rho$} is defined by
  \[    
    \Hq_{\alpha}^{\varepsilon}(\rho) 
  := 
    \frac{1}{1-\alpha} \log_2 \left(
      \inf_{\sigma \in \cB^{\varepsilon}(\rho)}
        \tr(\sigma^\alpha) \right) 
  \]
  with the convention $\Hq_{\alpha}^{\varepsilon}(\rho) := \lim_{\beta
    \to \alpha} \Hq_{\beta}^{\varepsilon}(\rho) $, for $\alpha=0$ or
  $\alpha=\infty$, and $\Hq_1^{\varepsilon}(\rho) := \Hq(\rho)$.
\end{definition}

The smooth R\'enyi entropy of order $\alpha$ can easily be expressed
in terms of conventional R\'enyi entropy. In particular, for $\alpha =
0$,
\begin{equation} \label{eq:Hzeroexp}
  \Hq_{0}^{\varepsilon}(\rho) 
= 
  \inf_{\sigma \in \cB^{\varepsilon}(\rho) } \Hq_0(\sigma)
\end{equation}
and, for $\alpha = \infty$, 
\begin{equation} \label{eq:Hinfexp}
  \Hq_{\infty}^{\varepsilon}(\rho) 
= 
  \sup_{\sigma \in \cB^{\varepsilon}(\rho) } \Hq_\infty(\sigma) \ .
\end{equation}

The following lemma is a direct generalization of the corresponding
classical statement in~\cite{RenWol03u}, saying that, for any order
$\alpha$, the smooth R\'enyi entropy $H_{\alpha}^{\varepsilon}(W)$ of
a random variable $W$ consisting of many independent and identically
distributed pieces asymptotically equals its Shannon entropy $H(W)$.

\begin{lemma} \label{lem:typsec}
  Let $\rho$ be a density operator. Then, for any $\alpha \in [0,
  \infty]$,
  \[
      \lim_{\varepsilon \to 0} \lim_{n \to \infty} 
        \frac{\Hq_{\alpha}^\varepsilon(\rho^{\otimes n})}{n}
    =
      \Hq(\rho) \ .
  \]  
\end{lemma}

\section{Secret keys and composability} \label{sec:uc}

The main idea for obtaining universally composable security
definitions is to compare the behavior of a \emph{real} cryptographic
protocol with an \emph{ideal} functionality. For a protocol which is
supposed to generate a secret key $S$, this ideal functionality is
simply a source which outputs an independent and uniformly distributed
random variable $U$ (in particular, $U$ is fully independent of the
adversary's information).  This motivates the following definition.

\begin{definition} \label{def:uc}
  Let $S$ be a random variable, let $\rvrho$ be a random state, and
  let $\varepsilon \geq 0$.  $S$ is said to be an
  \emph{$\varepsilon$-secure secret key with respect to $\rvrho$} if
  \[
    \dd(S|\rvrho) \leq \varepsilon \ .
  \]
\end{definition}

Consider a situation where $S$ is used as a secret key and where the
adversary's information is given by a random state $\rvrho$. The
$\varepsilon$-security of $S$ with respect to $\rvrho$ guarantees that
this situation (which is described by the density operator $[\rvrho
\otimes \{S\}]$) is $\varepsilon$-close---with respect to the trace
distance---to an ideal setting (described by $[\rvrho \otimes \{U\}]$)
where $S$ is replaced by a perfect key $U$ which is uniformly
distributed and independent of $\rvrho$.  Since the trace distance
does not increase when appending an additional quantum system
(cf.~(\ref{eq:trident})) or when applying any arbitrary quantum
operation (cf.~(\ref{eq:distop})), this also holds for any further
evolution of the system.  In particular, it follows
from~(\ref{eq:distmeas}) and Lemma~\ref{lem:vardistevent} that the
real and the ideal setting can be considered to be identical with
probability at least $1-\varepsilon$.

Definition~\ref{def:uc} is essentially equivalent to an intermediate
definition which has been used in~\cite{BHLMO02} to prove the
universal composability of QKD.  More precisely, if $S$ is
$\varepsilon$-secure according to Definition~\ref{def:uc}, it
satisfies the security definition of~\cite{BHLMO02} for some
$\varepsilon'$ depending on $\varepsilon$.\footnote{In~\cite{BHLMO02},
  a key $S$ about which an adversary has information $\rho_s$ is
  defined to be secure (with parameter $\varepsilon'$) if the Shannon
  distinguishability $\mathrm{SD}$ between $\rho_1:= \sum_s P_S(s)
  \proj{s} \otimes \rho_s$ and $\rho_0 := \sum_s \frac{1}{|\cS|}
  \proj{s} \otimes \rho'$, for $\rho' := \sum_{s} \frac{1}{|\cS|}
  \rho_s$, is small, i.e., $\varepsilon' = \mathrm{SD}(\rho_1,
  \rho_0)$. The relation between $\varepsilon$ and $\varepsilon'$ thus
  follows from the relation between the trace distance and the Shannon
  distance (see, e.g., \cite{FucGra99}).}  It is thus an immediate
consequence of the results in~\cite{BHLMO02} that
Definition~\ref{def:uc} provides universal composability in the
framework of~\cite{BenMay02}.

Note that Definition~\ref{def:uc} can also be seen as a natural
generalization of classical security definitions based on the
variational distance (which is the classical analogue of the trace
distance).  Indeed, if the adversary's knowledge is purely classical,
Definition~\ref{def:uc} is equivalent to the security definition as it
is, e.g., used in~\cite{DziMau04} or~\cite{KoMaRe03}.




\section{Main result} \label{sec:main}

\subsection{Theorem and proof}

\begin{theorem} \label{thm:hash}
  Let $Z$ be a random variable with range $\cZ$, let $\rvrho$ be a
  random state, and let $F$ be a \tu{} function on $\cZ$ with range
  $\cS = \{0,1\}^s$ which is independent of $Z$ and $\rvrho$. Then
  \[
    \dd(F(Z)|\{F\} \otimes \rvrho) 
  \leq 
     \frac{1}{2} 2^{-\frac{1}{2} (\Hq_2([\{Z\} \otimes \rvrho]) 
     - \Hq_0([\rvrho]) - s)} \ .
  \]
\end{theorem}

The following corollary is a consequence of
property~(\ref{eq:Halpharel}), expressions~(\ref{eq:Hzeroexp})
and~(\ref{eq:Hinfexp}), and the triangle inequality for the trace
distance.

\begin{corollary} \label{cor:eps}
  Let $Z$ be a random variable with range $\cZ$, let $\rvrho$ be a
  random state, let $F$ be a \tu{} function on $\cZ$ with range $\cS =
  \{0,1\}^s$ which is independent of $Z$ and $\rvrho$, and let
  $\varepsilon \geq 0$. Then
  \[
    \dd(F(Z)|\{F\} \otimes \rvrho) 
  \leq 
      \frac{1}{2}2^{-\frac{1}{2} 
        (\Hq_{\infty}^{\varepsilon}([\{Z\} \otimes \rvrho]) 
         - \Hq_0^{\varepsilon}([\rvrho]) - s)} 
         + 2 \varepsilon  \ .
  \]
\end{corollary}

Let us first state some technical lemmas to be used for the proof of
Theorem~\ref{thm:hash}.

\begin{lemma}\label{lem:basicdef}
  Let $Z$ be a random variable with range $\cZ$, let $\rvrho$ be a
  random state, and let $F$ be a random function with domain
  $\mathcal{Z}$ which is independent of $Z$ and $\rvrho$. Then
  \[
  \dd(F(Z)|\{F\}\otimes\rvrho) = E_F[\dd(F(Z)|\rvrho)].
  \]
\end{lemma}

\begin{proof}
  Let $U$ be a random variable uniformly distributed on $\cZ$ and
  independent of $F$ and $\rvrho$. Then
  \[
    \dd(F(Z)|\rvrho\otimes\{F\}) 
  =
  \dist\left(
    [(\{F(Z)\}\otimes\rvrho)\otimes\{F\}],
    [
    (\{U\}\otimes\rvrho)\otimes\{F\}
    ]
    \right),
  \]
  Now, applying Lemma~\ref{lem:distexp} to the random states
  $\{F(Z)\}\otimes\rvrho$ and $\{U\} \otimes \rvrho$ gives the desired
  result, since
  \[
  \begin{split}
    [\{F(Z)\}\otimes\rvrho|F=f] 
  & = 
    [\{f(Z)\}\otimes\rvrho]
  \\
    \left[\{U\}\otimes\rvrho|F=f\right]  
  & = 
    [\{U\}]\otimes[\rvrho] 
  \end{split}
  \]
  which holds because $F$ is independent of $Z$, $\rvrho$, and $U$.
\end{proof}

The following lemmas can most easily be formalized in terms of the
square of the Hilbert-Schmidt distance. For two density operators
$\rho$ and $\sigma$, let
\[
  \mydeltatwo(\rho, \sigma) 
:= 
  \tr \bigl((\rho - \sigma)^2 \bigr) \ .
\]
Moreover, for a random variable $X$ with range $\cX$ and a random
state $\rvrho$, we define
\[
  \mydtwo(X|\rvrho) 
:= 
  \mydeltatwo([\{X\} \otimes \rvrho], [\{U\}] \otimes [\rvrho]) 
\]
where $U$ is a random variable uniformly distributed on $\cX$.

\begin{lemma}\label{lem:tracedist}
  Let $\rho$ and $\sigma$ be two density operators on $\cH$. Then
  \[
    \dist(\rho,\sigma) 
  \leq 
    \frac{1}{2}\sqrt{\rank(\rho-\sigma) \cdot \mydeltatwo(\rho,\sigma)} \ .
  \]
\end{lemma}

\begin{proof}
  The assertion follows directly from Lemma~\ref{lem:absbound} and the
  definition of the distance measures $\dist(\cdot,\cdot)$ and
  $\mydeltatwo(\cdot,\cdot)$.
\end{proof}

\begin{lemma}\label{lem:donedtwo}
  Let $X$ be a random variable with range $\cX$ and let $\rvrho$ be a
  random state. Then
  \[
    \dd(X|\rvrho)
  \leq 
    \frac{1}{2} 2^{\frac{\Hq_0([\rvrho])}{2}}
      \sqrt{|\cX|\cdot\mydtwo(X|\rvrho)}.
  \]
\end{lemma}

\begin{proof}
  This is an immediate consequence of the definitions and
  Lemma~\ref{lem:tracedist}.
\end{proof}

\begin{lemma}\label{technicallemmaone}
  Let $X$ be a random variable with range $\cX$ and let $\rvrho$ be a
  random state. Then
  \[
    \mydtwo(X|\rvrho)
  =
    \tr\left(\bigl(\sum_{x\in \cX}P_X(x)^2\rho_x^2 \bigr)
      -\frac{1}{|\cX|}[\rvrho]^2\right)
  \]
  where $\rho_x:=[\rvrho|X=x]$ for $x \in \cX$.
\end{lemma}

\begin{proof}
  From~(\ref{eq:combexp}), we have
  \[ 
  \begin{split}
    \mydtwo(X|\rvrho) 
  & = 
    \tr \left(\left( \sum_{x\in\cX}P_X(x)\proj{x} \otimes\rho_x
            -\frac{1}{|\cX|}\sum_{x \in \cX}\proj{x} \otimes[\rvrho]\right)^2
    \right)\\
  & =
    \tr\left(\sum_{x \in \cX} 
      \bigl(P_X(x)\rho_x-\frac{1}{|\cX|}[\rvrho]\bigr)^2\right) \\
  & = \tr\left(
    \sum_{x\in\cX} P_X(x)^2\rho_x ^2 - \frac{2}{|\cX|}[\rvrho]
      \sum_{x\in\cX} P_X(x)\rho_x+\frac{1}{|\cX|}[\rvrho]^2.
    \right).
  \end{split}
  \]
  Inserting the identity
  \[
  [\rvrho]=\sum_{x\in\mathcal{X}}P_X(x)\rho_x
  \]
  concludes the proof.
\end{proof}

\begin{lemma}\label{technicallemmatwo}
  Let $Z$ be a random variable with range $\cZ$, let $\rvrho$ be a
  random state, and let $F$ be a \tu{} function on $\cZ$ chosen
  independently of $Z$ and $\rvrho$.  Then
  \[
    E_F\left[\mydtwo(F(Z)|\rvrho)\right]
  \leq
    2^{-\Hq_2([\{Z\}\otimes\rvrho])} \ .
  \]
\end{lemma}

\begin{proof} 
  Let us define $\rho_z:=[\rvrho|Z=z]$ for every $z \in \cZ$ and let
  $\cS$ be the range of $F$.  With Lemma~\ref{technicallemmaone}, we
  obtain
  \begin{equation} \label{eq:myexpressionone}
  \begin{split}
    E_F\left[\mydtwo(F(Z)|\rvrho)\right] 
  = 
    \tr\left(
      E_F\left[\sum_{s\in\mathcal{S}} 
        \Pr[F(Z)=s]^2[\rvrho|F(Z)=s]^2\right]\right)-\frac{1}{|\cS|}
        \tr([\rvrho]^2) \ ,
  \end{split}
  \end{equation}
  using the linearity of the expectation value and the trace.  Note
  that
  \[
    {\Pr[f(Z)=s]}\cdot[\rvrho|f(Z)=s] \,
  =
    \sum_{z\in f^{-1}(\{s\})} P_Z(z)\rho_z \ .
  \]
  Using this identity and rearranging the summation order, we get
  \[
    \sum_{s \in \cS} \Pr[f(Z)=s]^2[\rvrho|f(Z)=s]^2
  = 
    \sum_{z,z'\in\mathcal{Z}}
      P_Z(z)P_Z(z')\rho_z\rho_{z'}\delta_{f(z),f(z')} \ ,
  \]
  where $\delta_{x,y}$ is the Kronecker delta which equals $1$ if
  $x=y$ and $0$ otherwise.  Taking the expectation value over the
  random choice of $F$ then gives
  \[
    E_F\left[\sum_{s\in\mathcal{S}}\Pr[F(Z)=s]^2[\rvrho|F(Z)=s]^2\right]
  =
    \sum_{z,z'\in\mathcal{Z}} P_Z(z)P_Z(z')\rho_z\rho_{z'} \Pr[F(z)=F(z')] \ .
  \]
  Similarly, we obtain
  \[
    [\rvrho]^2
  =
    \sum_{z,z'\in\mathcal{Z}}P_Z(z)P_Z(z')\rho_z\rho_{z'} \ .
  \]
  Inserting this into~(\ref{eq:myexpressionone}), we get
  \[
    E_F\left[\mydtwo(F(Z)|\rvrho)\right]
  = 
    \sum_{z,z'\in\mathcal{Z}}
      P_Z(z)P_Z(z') 
        \left( \Pr[F(z)=F(z')]-\frac{1}{|\mathcal{S}|} \right)
        \tr(\rho_z\rho_{z'}) \ .
  \]
  As we assumed that $F$ is \tu{}, all summands with $z\neq z'$ are
  not larger than zero and we are left with
  \[
    E_F\left[\mydtwo(F(Z)|\rvrho)\right]
  \leq
    \sum_{z\in\mathcal{Z}} P_Z(z)^2 \tr(\rho_z^2) 
  =
    \tr \bigl([\{Z\} \otimes \rvrho]^2 \bigr)
  \]
  from which the assertion follows by the definition of the R\'enyi
  entropy $\Hq_2$.
\end{proof}

\begin{proof}[Proof of Theorem~\ref{thm:hash}]
  Using Lemma~\ref{lem:basicdef}, Lemma~\ref{lem:donedtwo}, we get
  \[
  \begin{split}
    \dd(F(Z)|\{F\}\otimes\rvrho)
  & = 
    E_F[\dd(F(Z)|\rvrho)] \\
  & \leq 
   \frac{1}{2} 
     2^{\frac{s+\Hq_0([\rvrho])}{2}} E_F[\sqrt{\mydtwo(F(Z)|\rvrho)}] \\
  & \leq 
    \frac{1}{2} 
      2^{\frac{s+\Hq_0([\rvrho])}{2}} \sqrt{E_F[\mydtwo(F(Z)|\rvrho)]} \ .
  \end{split}
  \]  
  where the last inequality follows from Jensen's inequality and the
  convexity of the square root. Applying Lemma~\ref{technicallemmatwo}
  concludes the proof.
\end{proof}

\subsection{Privacy amplification against quantum adversaries} \label{sec:appl}


We now apply the results of the previous section to show that privacy
amplification by \tu{} hashing is secure (with respect to the
universally composable security definition of Section~\ref{sec:uc})
against an adversary holding quantum information.  Consider two
distant parties which are connected by an authentic, but otherwise
fully insecure classical communication channel.  Additionally, they
have access to a common random string $Z$ about which an adversary has
some partial information represented by the state $\rvrho$ of a
quantum system. The two legitimate parties can apply the following
\emph{privacy amplification protocol} to obtain a secure key $S$: One
of the parties chooses an instance of a \tu{} function $F$ and
announces his choice to the other party using the public communication
channel.  Then, both parties compute $S=F(Z)$.  Since the information
of the adversary after the execution of the protocol is given by
$\rvrho \otimes \{F\}$, one wants the final key $S$ to be
$\varepsilon$-secure with respect to $\rvrho \otimes \{F\}$ (cf.\ 
Definition~\ref{def:uc}), for some small $\varepsilon \geq 0$.  It is
an immediate consequence of Corollary~\ref{cor:eps} that this is
achieved if the key $S$ has length at most
\begin{equation} \label{eq:pa}
  s 
= 
  \Hq_{\infty}^{\epsb}([\{Z\} \otimes \rvrho]) 
  - \Hq_0^{\epsb}([\rvrho]) 
  - 2 \log_2(\frac{1}{4\epsb}) \ ,
\end{equation}
for $\epsb = \varepsilon/4$.

\subsection{Asymptotic optimality} \label{sec:opt}

We now show that the bound~(\ref{eq:pa}) is asymptotically optimal,
i.e., that the right hand side of~(\ref{eq:pa}) is (in an asymptotic
sense) also an upper bound for the number of key bits that can be
extracted by any protocol.  Consider a setting where both the initial
information $Z^{(n)}$ as well as the adversary's state $\rvrho^{(n)}$
consist of many independent pieces: For $n \in \bbN$, let $Z^{(n)} =
(Z_1, \ldots, Z_n)$ and $\rvrho^{(n)} = \rvrho_1 \otimes \cdots
\otimes \rvrho_n$ where $(Z_i, \rvrho_i)$ are independent pairs with
identical probability distribution $P_{(Z_i, \rvrho_i)} = P_{(Z,
  \rvrho)}$. Let $s(n)$ be the length of the key $S$ that can be
extracted from $Z^{(n)}$ by an optimal privacy amplification protocol.
Using Lemma~\ref{lem:typsec}, we conclude from~(\ref{eq:pa}) that
\begin{equation} \label{eq:lower}
  s(n) \geq H(Z^{(n)}|\rvrho^{(n)}) + o(n)
\end{equation}
where, for any $\bar{Z}$ and $\bar{\rvrho}$, $H(\bar{Z}|\bar{\rvrho})$
is defined by
\[
  H(\bar{Z}|\bar{\rvrho})
:= 
  \Hq([\{\bar{Z}\} \otimes \bar{\rvrho}]) - \Hq([\bar{\rvrho}]) \ .
\]
Let now $S:=F(Z^{(n)})$ be a key of length $s(n)$ computed by applying
any random function $F$ to $Z^{(n)}$. It is a direct consequence of
Definition~\ref{def:uc} that the key $S$ can only be
$\varepsilon$-secure with respect to $\rvrho^{(n)} \otimes \{F\}$ (for
$\varepsilon$ approaching $0$ as $n$ goes to infinity) if
\begin{equation} \label{eq:upper}
  s(n) \leq H(F(Z^{(n)})|\rvrho^{(n)} \otimes \{F\}) + o(n) \ .
\end{equation}
Note that the quantity $H(\bar{Z}|\bar{\rvrho})$ can only decrease
when applying any function $f$ to its first argument, i.e., for any
random function $F$ chosen independently of $Z^{(n)}$ and $\rvrho$,
\begin{equation} \label{eq:monotone}
  H(F(Z^{(n)})|\rvrho^{(n)} \otimes \{F\}) 
\leq 
  H(Z^{(n)}|\rvrho^{(n)} \otimes \{F\})
= 
  H(Z^{(n)}|\rvrho^{(n)}) \ .
\end{equation}
Thus, combining~(\ref{eq:lower}), (\ref{eq:upper}),
and~(\ref{eq:monotone}), we obtain an expression for the maximum
number $s(n)$ of extractable key bits,
\[
  s(n) = H(Z^{(n)}|\rvrho^{(n)}) + o(n) \ .
\]
In particular, the maximum rate $R := \lim_{n \to \infty}
\frac{s(n)}{n}$ at which secret key bits can be generated, from
independent realizations of $Z$ about which the adversary has
information given by $\rvrho$, is
\begin{equation} \label{eq:rate}
  R  = \Hq([\{Z\} \otimes \rvrho]) - \Hq([\rvrho]) = H(Z | \rvrho) \ .
\end{equation}

This fact is already known for the special case where the adversary's
information is purely classical. Indeed, if the adversary's knowledge
about each realization of $Z$ is given by a realization of a random
variable $W$, expression~(\ref{eq:rate}) reduces to the well-known
classical result
\[
  R = H(Z W) - H(W) = H(Z|W) 
\]
(see, e.g., \cite{CsiKor78} or \cite{Maurer93}).

\subsection{Applications to QKD} \label{sec:QKD}

Theorem~\ref{thm:hash} has interesting implications for quantum key
distribution (QKD). Recently, a generic protocol for QKD has been
presented and proven secure against general attacks~\cite{ChReEk04}.
Moreover, it has been shown that many of the known protocols, such as
BB84 or B92, are special instances of this generic protocol, i.e.,
their security directly follows from the security of the generic QKD
protocol. Since the result in~\cite{ChReEk04} is based on the security
of privacy amplification, the strong type of security implied by
Theorem~\ref{thm:hash} immediately carries over to this generic QKD
protocol.  In particular, the secret keys generated by the BB84 and
the B92 protocol satisfy Definition~\ref{def:uc} and thus provide
universal composability.

\section{Acknowledgment}

The authors thank Ueli Maurer for many inspiring discussions, and
Dominic Mayers for useful comments.

\appendix

\section{Some identities}

\begin{lemma}[Schur's inequality] \label{lem:schur}
  Let $A$ be a linear operator on a $d$-dimensional Hilbert space
  $\cH$ and let $\lambda_1, \ldots, \lambda_d$ be its eigenvalues.
  Then
  \[
      \sum_{i=1}^d |\lambda_{i}|^2
    \leq
      \tr(A A^{\dagger}) \ ,
  \]
  with equality if and only if $A$ is normal (i.e., $A
  A^{\dagger}=A^{\dagger} A$).
\end{lemma}

\begin{proof}
  See, e.g., \cite{HorJoh85}.
\end{proof}
 

\begin{lemma} \label{lem:absbound}
  Let $A$ be a normal operator with rank $r$. Then
  \[
    \tr |A| \leq \sqrt{r} \sqrt{\tr (A A^{\dagger})} \ .
  \]
\end{lemma}

\begin{proof}
  Let $\lambda_1, \ldots, \lambda_r$ be the $r$ nonzero eigenvalues of
  $A$.  Since the square root is concave, we can apply Jensen's
  inequality leading to
  \[
    \tr|A| 
  = 
    \sum_{i=1}^{r} |\lambda_i|
  =
    \sum_{i=1}^{r} \sqrt{|\lambda_i|^2}
  \leq 
    \sqrt{r} \sqrt{\sum_{i=1}^{r} |\lambda_i|^2} \ .
  \]
  The assertion then follows from Schur's inequality. 
\end{proof}



\begin{thebibliography}{10}

\bibitem{BeBrRo88}
C.~H. Bennett, G.~Brassard, and J.-M. Robert.
\newblock Privacy amplification by public discussion.
\newblock {\em SIAM Journal on Computing}, 17(2):210--229, 1988.

\bibitem{ImLelu89}
R.~Impagliazzo, L.~A. Levin, and M.~Luby.
\newblock Pseudo-random generation from one-way functions (extended abstract).
\newblock In {\em Proceedings of the Twenty-First Annual ACM Symposium on
  Theory of Computing}, pages 12--24, 1989.

\bibitem{BBCM95}
C.~H. Bennett, G.~Brassard, C.~Cr{\'e}peau, and U.~Maurer.
\newblock Generalized privacy amplification.
\newblock {\em IEEE Transaction on Information Theory}, 41(6):1915--1923, 1995.

\bibitem{KoMaRe03}
R.~K\"onig, U.~Maurer, and R.~Renner.
\newblock On the power of quantum memory.
\newblock Available at {\tt http://arxiv.org/abs/quant-ph/0305154}, 2003.

\bibitem{ChReEk04}
M.~Christandl, R.~Renner, and A.~Ekert.
\newblock A generic security proof for quantum key distribution.
\newblock Available at {\tt http://arxiv.org/abs/quant-ph/0402131}, February
  2004.

\bibitem{Canetti01}
R.~Canetti.
\newblock Universally composable security: A new paradigm for cryptographic
  protocols.
\newblock In {\em Proc.\ 42nd IEEE Symposium on Foundations of Computer Science
  (FOCS)}, pages 136--145, 2001.

\bibitem{PfiWai00a}
B.~Pfitzmann and M.~Waidner.
\newblock Composition and integrity preservation of secure reactive systems.
\newblock In {\em 7th ACM Conference on Computer and Communications Security},
  pages 245--254. ACM Press, 2000.

\bibitem{BenMay02}
M.~Ben-Or and D.~Mayers.
\newblock Quantum universal composability.
\newblock Slides available at
  {\tt http://www.msri.org/publications/ln/msri/2002/quantumcrypto/mayers/1}, 2002.

\bibitem{BHLMO02}
M.~Ben-Or, M.~Horodecki, D.~Leung, D.~Mayers, and J.~Oppenheim.
\newblock Composability of {QKD}.
\newblock Talk given by D.~Mayers. Slides available at
  {\tt http://www.msri.org/publications/ln/msri/2002/qip/mayers/1} (Part~II),
  2002.

\bibitem{NieChu00}
M.~A. Nielsen and I.~L. Chuang.
\newblock {\em Quantum computation and quantum information}.
\newblock Cambridge University Press, 2000.

\bibitem{GotLo03}
D.~Gottesman and H.-K. Lo.
\newblock Proof of security of quantum key distribution with two-way classical
  communications.
\newblock {\em IEEE Transactions on Information Theory}, 49(2):457--475, 2003.

\bibitem{BenBra84}
C.~H. Bennett and G.~Brassard.
\newblock Quantum cryptography: Public-key distribution and coin tossing.
\newblock In {\em Proceedings of IEEE International Conference on Computers,
  Systems and Signal Processing}, pages 175--179, 1984.

\bibitem{Bennett92}
C.~H. Bennett.
\newblock Quantum cryptography using any two nonorthogonal states.
\newblock {\em Physical Review Letters}, 68(21):3121--3124, 1992.

\bibitem{CarWeg79}
J.~L. Carter and M.~N. Wegman.
\newblock Universal classes of hash functions.
\newblock {\em Journal of Computer and System Sciences}, 18:143--154, 1979.

\bibitem{WegCar81}
M.~N. Wegman and J.~L. Carter.
\newblock New hash functions and their use in authentication and set equality.
\newblock {\em Journal of Computer and System Sciences}, 22:265--279, 1981.

\bibitem{Renyi61}
A.~R\'enyi.
\newblock On measures of entropy and information.
\newblock In {\em Proceedings of the 4th Berkeley Symp.\ on Math.\ Statistics
  and Prob.}, volume~1, pages 547--561. Univ.\ of Calif.\ Press, 1961.

\bibitem{RenWol03u}
R.~Renner and S.~Wolf.
\newblock Smooth {R\'enyi} entropy and applications.
\newblock Accepted for ISIT 2004. Available at
  {\tt http://www.crypto.ethz.ch/\~{}renner/publications.html}, October 2003.

\bibitem{FucGra99}
C.~A. Fuchs and J.~van~de Graaf.
\newblock Cryptographic distinguishability measures for quantum mechanical
  states.
\newblock {\em IEEE Transactions on Information Theory}, 45(4):1216--1227,
  1999.
\newblock Available at {\tt http://arxiv.org/abs/quant-ph/9712042}.

\bibitem{DziMau04}
S.~Dziembowski and U.~Maurer.
\newblock Optimal randomizer efficiency in the bounded-storage model.
\newblock {\em Journal of Cryptology}, 17(1):5--26, 2004.
\newblock Conference version appeared in Proc.\ of \mbox{STOC '02}.

\bibitem{CsiKor78}
I.~Csisz\'{a}r and J.~K\"{o}rner.
\newblock Broadcast channels with confidential messages.
\newblock {\em IEEE Transactions on Information Theory}, 24:339--348, 1978.

\bibitem{Maurer93}
U.~M. Maurer.
\newblock Secret key agreement by public discussion from common information.
\newblock {\em IEEE Transactions on Information Theory}, 39(3):733--742, 1993.

\bibitem{HorJoh85}
R.~A. Horn and C.~R. Johnson.
\newblock {\em Matrix analysis}.
\newblock Cambridge University Press, 1985.

\end{thebibliography}

\end{document}